\documentclass{article}

\usepackage{arxiv}
\usepackage[utf8]{inputenc} 
\usepackage[T1]{fontenc}    
\usepackage{hyperref}       
\usepackage{url}            
\usepackage{booktabs}       
\usepackage{amsfonts}       
\usepackage{nicefrac}       
\usepackage{microtype}      
\usepackage{lipsum}
\usepackage{amsmath, xcolor, color, soul}
\usepackage{appendix}
\usepackage{graphicx}
\usepackage{multirow}
\usepackage{indentfirst}
\usepackage{caption}
\graphicspath{ {./images/} }
\usepackage{cite}
\title{SuperCond-GNN: Scalable Graph Neural Network Surrogate for Superconducting Circuit Simulations}

\author{
Nandana Menon \\
Acclerator Technology \& Applied Physics Division\\
  Lawrence Berkeley National Laboratory\\ Berkeley, CA\\
  \texttt{nmenon@lbl.gov} \\
   \And
Giorgio Vallone \\
Engineering Division\\
  Lawrence Berkeley National Laboratory\\ Berkeley, CA\\
  \texttt{gvallone@lbl.gov} \\}

\begin{document}
\maketitle
\begin{abstract}
This paper presents SuperCond-GNN, a graph neural network-based surrogate model 
for predicting the voltage distribution in high-temperature superconducting (HTS) 
magnets. HTS magnets are modeled as lumped-element 
equivalent circuits and mapped onto graph representations, enabling message 
passing GNNs to learn the electrical response as a function of circuit topology, 
material properties, and operating current. As a proof of concept, tape stacks of up to 10 tapes are considered across a 
range of circuit topologies and operating conditions. The surrogate is trained on data generated from circuit simulations and achieves 
a mean MAPE of $4.3$\% within the prescribed design space. The predicted nodal 
voltages enable fast and scalable inference of current redistribution and local 
operating conditions across a wide range of circuit configurations. The effect 
of incorporating physics-informed regularization via Kirchhoff's current law is also 
evaluated, and generalizability to unseen topologies is assessed through zero-shot 
inference and few-shot fine-tuning. While demonstrated on tape stack circuits, 
the graph-based framework is topology-agnostic and naturally extensible to more 
complex HTS cable and magnet configurations, offering a scalable alternative to 
conventional circuit solvers for downstream applications such as design space 
exploration, current sharing analysis, and real-time magnet monitoring.

\end{abstract}


\section{Introduction}

High-temperature superconducting (HTS) magnets based on REBCO conductors offer unprecedented performance for accelerator, fusion, and high-field applications, but protecting them remains a major challenge. Because slow normal-zone propagation and high heat capacity severely limit the effectiveness of conventional protection strategies{~\cite{marchevsky_quench_2021}}, voltage-based detection systems often identify a quench only after it has already occurred. This delay leaves magnets vulnerable to damage from localized heating and high voltages. Additional complications arise current sharing in cable and coil systems which is governed by a complex interplay of termination resistance, contact resistance, conductor non-uniformity, local defects, magnet history, and architecture-dependent redistribution pathways. Together, these factors can delay quench onset and make global voltage-current characteristics exceedingly difficult to interpret at the global magnet level.

{Transitioning from reactive to predictive protection requires both improved sensing and real-time models capable of capturing the magnet's state to anticipate failure. This need motivates the broader goal of developing an AI-enabled, physics-informed digital twin{~\cite{rasheed_digital_2020}} for the diagnostics, protection, and control of HTS magnet systems}. A critical first step toward this framework is the ability to reproduce relevant electromagnetic behavior at the cable or subscale level with high fidelity and computational efficiency. Currently, lumped-element circuit models implemented in SPICE simulators are a standard tool for understanding and predicting these behaviors in both steady-state~\cite{martinez_electric-circuit_2020, kang_current_2023, pothavajhala_experimental_2014} and transient~\cite{willering_effect_2015, berger_stability_2011} regimes. Given accurate parameter measurements, they can reliably predict real device behavior~\cite{pothavajhala_experimental_2014, willering_effect_2015}. Yet, scaling these models from individual tapes to cables and full magnets drastically increases circuit complexity and simulation cost. Parametric studies compound this issue by requiring thousands of individual simulations. Consequently, SPICE-based models lack the scalability needed to serve as real-time estimators in a digital twin environment. Furthermore, GPU-accelerated implementations like CUSPICE~\cite{lannutti_cuspice_2014} are restricted to linear circuit components, making them insufficient for accurately representing superconducting circuits.

A widely adopted strategy for accelerating simulations and augmenting experiments is the use of machine-learning-based (ML) surrogates, an approach that has proven highly effective across numerous domains. In the context of HTS magnets, existing work has focused primarily on forecasting models for quench detection from experimental sensor signals~\cite{sakakibara_experimental_2026, khan_weakly_2026}. Despite these advances, relatively little work has addressed true physics surrogates capable of real-time deployment within a broader digital twin framework. Xiao et al.~\cite{xiao_surrogate_2026} represent the closest effort in this direction, developing an FEM surrogate to predict current density distributions in REBCO solenoids. However, conventional architectures such as feedforward neural networks employed in \cite{xiao_surrogate_2026} operate on fixed-dimensional input spaces, requiring the entire magnet or its circuit representation to be encoded into a vector of predetermined size. Consequently, such surrogate models are inherently design-specific, trained for 
a fixed magnet topology and capable only of interpolating within the parameter 
space seen during training. From a circuit model perspective, this is a fundamental limitation since the number of components varies with design and hence raises a natural question: how should the circuit be represented such that a surrogate can generalize across variable-size configurations and learn the underlying physical behavior?

\begin{figure}
    \centering
    \includegraphics[width=1\linewidth]{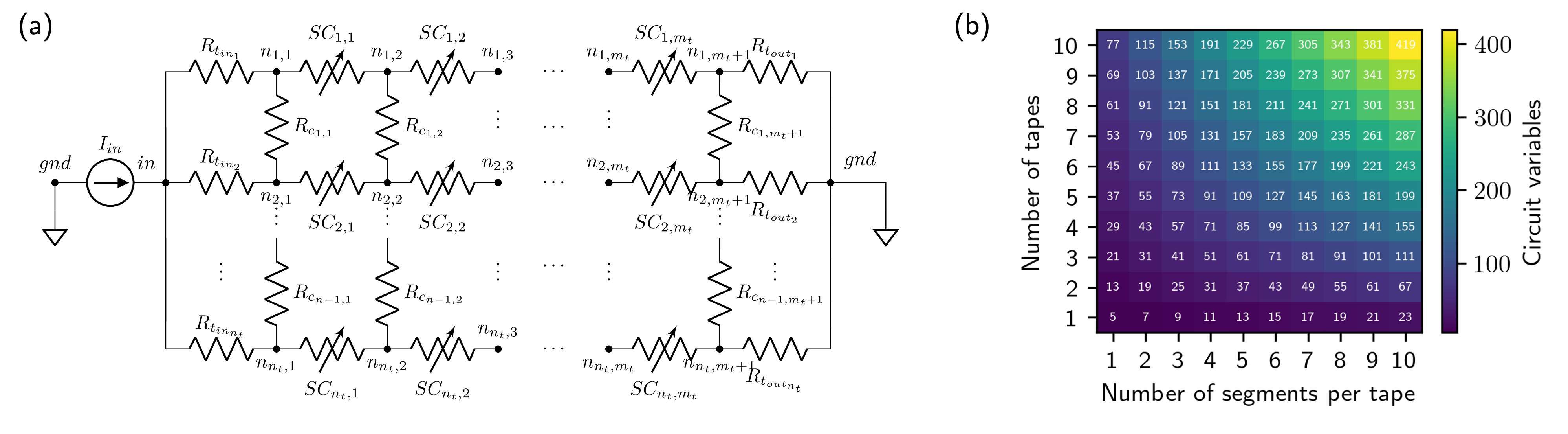}
    \caption{(a) Circuit model for tape stack cable with $n_t$ REBCO tapes, each divided into $m_t$ segments. The contact resitances between the tapes are denoted by $R_{c_{i,j}}$. $R_{{t_{in}}_i}$ and $R_{{t_{out}}_i}$ are the termination resistances. (b) Heatmap showing the number of variables as $n_t$ and $m_t$ are varied.}
    \label{fig:circuit_intro}
\end{figure}

This limitation becomes evident even in fundamental scenarios, such as simulating the transport measurement of a single REBCO tape. Assuming a highly simplified, lumped-parameter model, the necessary inputs to predict the measured voltage ($V$) are restricted to the applied transport current, terminal resistances ($R_t$), critical current ($I_c$), and power-law exponent ($n$). While a conventional neural network can be trained on a design-of-experiments dataset to map this fixed input space to the resulting voltage, its rigid architecture fundamentally breaks down when scaling. Extending this model to a stack of two tapes alters the required input dimensionality by introducing an entirely new set of $I_c$ and $n$ values, alongside contact resistances ($R_c$) that dictate current sharing between the tapes. Furthermore, achieving high-fidelity simulations requires discretizing the tapes into segments to capture localized spatial variations in $I_c$ and $n$, as well as the distributed influence of their contact. Figure~\ref{fig:circuit_intro}(a) illustrates the resulting electric circuit model for an $n_t$-tape stack divided into $m_t$ segments. As shown in Figure~\ref{fig:circuit_intro}(b), the total parameter space grows rapidly with the addition of tapes and segments, even when restricted to steady-state circuit parameters ($R_t$, $R_c$, transport current $I_{in}$, $I_c$, and $n$). This dimensional scaling becomes even more severe for complex geometries, such as CORC cables, which necessitate the inclusion of winding pitch and tape radius~\cite{teyber_numerical_2022, willering_effect_2015}. Consequently, employing a conventional neural network would require either imposing an arbitrary maximum circuit size to fix the input dimensionality, or training a separate model for every distinct geometry. Neither approach is practical or generalizable. Additionally, omitting parameters to artificially enforce a fixed input size is physically unsound, as nodal voltages depend on the global interaction of all circuit elements. Ultimately, the inherently variable dimensionality of superconducting circuit models renders standard neural networks unsuitable as generalizable surrogates for these systems.

Graph neural networks (GNNs) overcome this limitation by operating directly on 
graph-structured data composed of nodes and edges. Through permutation-invariant 
aggregation over node and edge feature spaces, GNNs can process graphs of 
arbitrary size without architectural modification, requiring only that the node 
and edge features lie within a fixed feature space irrespective of the number 
of nodes or edges. By employing message passing mechanisms to learn relationships 
between components, GNNs enable predictions at the node, edge, or graph level, 
making them a natural fit for a truly generalizable circuit surrogate. GNNs have been broadly adopted across diverse domains, including drug discovery, recommendation systems, traffic forecasting, and materials property prediction. In high-energy physics, GNNs have seen widespread use in charged-particle tracking, real-time data filtering and compression, and particle-flow reconstruction~\cite{dezoort_graph_2023}. In electronics, their use has been comparatively limited, with most applications concentrated in digital and analog circuit design. More recently, GNNs have been applied to graph-level regression for predicting circuit performance metrics\cite{khamis_circuit_2024}. Most pertinent to the present work is the study by Hakhamaneshi et al.~\cite{hakhamaneshi_pretraining_2022}, who employed a deep graph convolutional network to predict DC nodal voltages in resistor-based and operational-amplifier circuits. Beyond demonstrating node-level prediction accuracy, their work showed that GNN generalizability extends to few-shot learning scenarios where a model pretrained on smaller circuits could be successfully transferred to larger, unseen topologies through fine-tuning on a substantially reduced dataset. This finding is particularly relevant to the present work, where generalization across variable circuit sizes is a primary objective.

This work presents SuperCond-GNN (SC-GNN), a graph neural network (GNN)-based surrogate model for predicting the steady-state behavior of HTS magnets and cables represented as lumped-element circuit models. By representing the circuit as a graph, SC-GNN naturally accommodates variable topologies and sizes without retraining or architectural modification, addressing a fundamental limitation of conventional surrogate approaches. As a proof of concept, REBCO tape stack cables are considered, which capture the essential current-sharing and nonlinear resistive behavior characteristic of practical HTS systems. The target quantity of interest is the nodal voltage distribution, which is the natural output of circuit simulations as well as measurements. Predicting these voltages is directly relevant to voltage-based quench detection systems, representing a key step toward the ultimate goal of integrating this surrogate within a real-time digital twin framework for HTS magnet protection. The key contributions of this work are: (i) a graph-based circuit representation for HTS tape stacks enabling topology-agnostic surrogate modeling, (ii) a physics-informed GNN trained with Kirchhoff's current law (KCL) regularization, (iii) comprehensive benchmarking against ngspice across the full topology space, and (iv) an assessment of zero-shot and few-shot generalizability to out-of-distribution circuit configurations.

\section{Methodology}
\label{sec:headings}
\subsection{Electric-circuit model of tape stack cable}

The circuit model employed in this work is the 2D network model developed by Martínez et al.~\cite{martinez_electric-circuit_2020} for steady-state simulation of a tape stack cable consisting of $n_t$ REBCO tapes. Each tape is divided into $m_t$ segments, such that the conductor is represented as a serial chain of current-dependent voltage sources governed by the power law relation describing the current-voltage characteristics of an HTS conductor:
\begin{equation}
V_{i,j}(I_{i,j}) = V_{c}\left(\frac{I_{i,j}}{I_{c_{i,j}}}\right)^{n_{i,j}}
\end{equation}
where $i$ and $j$ denote the tape and segment indices, respectively, $I_{i,j}$ is the current flowing in that element, $I_{c_{i,j}}$ is the critical current, and $n_{i,j}$ the $n$-value for that element. The criterion voltage $V_{c}$ is derived from the standard electric field criterion of $1\,\mu\text{V/cm}$, applied under the assumption that each tape has a total length of 10\,cm divided into equal segments. All tapes within the stack are assumed to be of equal length. Therefore, each element $SC_{i,j}$ in Figure.~\ref{fig:circuit_intro}(a) represents the aforementioned voltage source. Circuits are simulated using DC analysis in ngspice~\cite{vogt_ngspice_nodate}, an open source SPICE simulator, with no time dependence. The transport current is ramped to a target value, $I_\text{target} = 0.8 \times n_t I_c$, that captures the transition of the stack while ensuring numerical convergence of the nonlinear solver~\cite{yu_experimental_2024}. Nodal voltages are then extracted from the solver to produce $V-I$ curves characteristic of short sample tests of such conductors.

\subsection{Graph representation of circuits for GNNs}

While circuits are natural graph representations, for the purpose of adapting them to GNNs, selecting an appropriate form graph representation is critical. In the context of GNNs, a graph, $G$, is defined by a set of nodes, $V$, and a set of edges, $E$, encoding the pairwise relationships between them, i.e., $G=(V,E)$. Each node is associated with a feature vector while edges may or may not be. Unlike more established application domains, no unified convention currently exists for encoding electrical circuits as graphs for GNN. The most intuitive and physically grounded approach follows classical graph theory, in which circuit nodes map to graph nodes and circuit elements map to edges\cite{mo_graph_2022}, and this convention is adopted here in preference to alternative representations\cite{khamis_circuit_2024, hakhamaneshi_pretraining_2022}. Each node and edge is then assigned a set of features encoding its physical and topological information.

Two binary node features are defined to encode boundary conditions which identifies whether a node serves as the ground reference, or the terminal at which the external current is injected. Together, these features identify the boundary points of the circuit. The ground node also serves as a node whose voltage is known \textit{a priori}. To encode topological information, three additional features are introduced: normalized degree of the node, normalized distance from the ground and normalized distance from the current source. The normalized degree captures the local connectivity of a node, while the latter two features convey global structural information by quantifying the node's distance from the ground node and the current-injection node, respectively, thereby characterizing the overall current-flow path through the circuit. These distances are computed via breadth-first search (traversing along segments within a tape rather than across tapes) originating from the ground node or the current-injection node and are subsequently normalized by the maximum distance observed in the graph.

\begin{table}[]
\caption{Summary of node and edge features used in the graph representation 
of the HTS tape stack circuit model. Categorical features are binary indicators 
encoded as one-hot vectors, while numeric features are continuous physical 
quantities passed directly as scalars.}
\centering
\label{tab:node-edge}
\renewcommand{\arraystretch}{1.2}
\begin{tabular}{l|l|l|l|l}
\hline
Index & Node feature                       & Type        & Edge Features         & Type        \\ \hline
1     & Ground node                        & Categorical & Resistor              & Categorical \\
2     & Input node                         & Categorical & Superconductor        & Categorical \\
3     & Normalized node degree             & Numeric     & Current source        & Categorical \\
4     & Normalized ground distance         & Numeric     & $R$                   & Numeric     \\
5     & Normalized current source distance & Numeric     & $I_c$                 & Numeric     \\
6     & --                                  & --           & $n$                   & Numeric     \\
7     & --                                  & --         & $I_{in}$     & Numeric     \\
8     & --                                  & --           & $I_{in}/I_c$ & Numeric    \\ \hline
\end{tabular}
\end{table}

\begin{figure}
    \centering
    \includegraphics[width=1\linewidth]{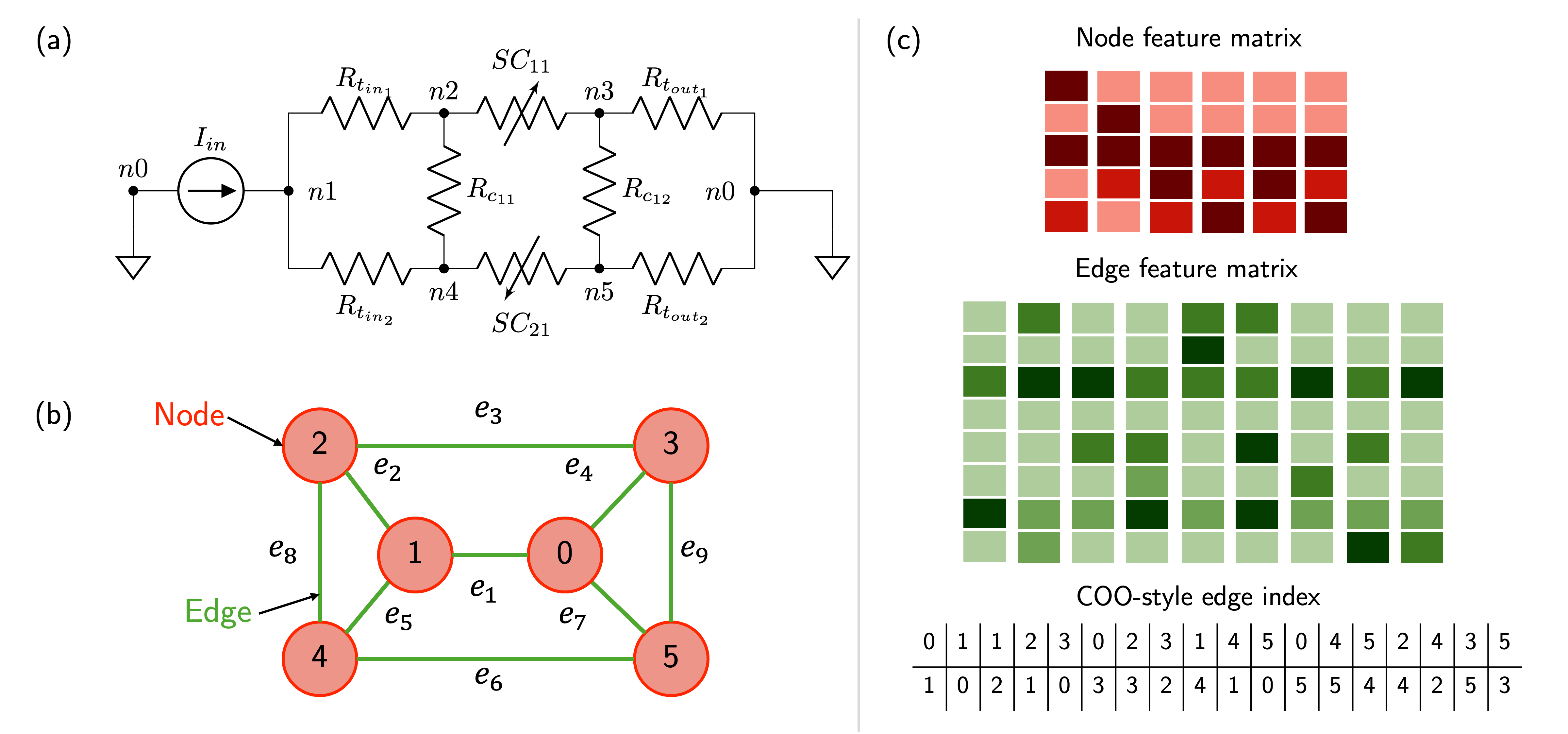}
    \caption{Graph representation of a tape stack circuit model with two tapes without any discretizations. (a) Schematic of a $2\times1$ tape stack, comprising terminal resistances $R_{t_{in}}$ and $R_{t_{out}}$, superconducting elements $SC_{11}$ and $SC_{21}$, and contact resistances $R_{c_{11}}$ and $R_{c_{12}}$, driven by a transport current $I_{in}$. (b) Corresponding graph representations, where circuit nodes (red) and elements (green edges) map directly to graph nodes and edges respectively. (c) The associated node feature matrix, edge feature matrix, and COO-style edge index, where each edge appears twice to reflect the undirected connectivity of the graph.}
    \label{fig:graph-intro}
\end{figure}

Edge features encode the physical properties of the components in the circuit. These features must capture both the element type and its associated electrical parameters. The circuit elements are one-hot encoded (binary vector representation in which a single entry is set to one to indicate the element type and all others are set to zero) to identify if it is a resistor, superconducting element or the current source. Depending on the element type, the corresponding physical parameters are then populated. For resistive elements, which include contact resistances and terminal resistances, the associated resistance value is recorded. For the current source, the input current magnitude $I_{in}$ is recorded. For superconducting elements, the critical current $I_c$, the power-law exponent $n$, and the normalized operating ratio $I_{in}/I_c$ are specified, where the latter provides a dimensionless measure of how close the element is to its critical current and thus captures its nonlinear operating regime. All unused parameter fields for a given element type are set to zero. Each edge is, therefore, described by an eight-dimensional feature vector jointly encoding the element type and its governing physical parameters. The features are summarized in Table.~\ref{tab:node-edge}. Figure~\ref{fig:graph-intro} illustrates the graph representation of a circuit, where panel (a) shows the schematic and panel (b) shows the corresponding graph. Panel (c) shows the node and edge feature matrices of the graph. Connectivity is defined using a Coordinate List (COO) format, where each edge is represented as a pair of node indices. Since edges are undirected to reflect the mutual relationship between circuit components, each coordinate pair appears twice in the COO table,  once for each direction. An ablation study justifying the choice 
of input features is provided in Appendix A.

\subsection{Model architecture}

The surrogate model in this study, SuperCond-GNN (SC-GNN), is based on the MeshGraphNet architecture~\cite{pfaff_learning_2021} originally developed for learning mesh-based simulations. The architecture consists of three stages: an encoder that projects raw node and edge features into a shared latent space, a processor that performs iterative message passing to propagate information across the graph, and a decoder that maps the updated latent node representations to the target output, in this case, the nodal voltages.

In the encoding stage, the node feature vector and edge feature vector are each independently projected into a common hidden dimension $d_h$  through dedicated two-layer multilayer perceptrons (MLPs) with ReLU activation and layer normalization. This shared dimensionality enables the subsequent processor layers to operate uniformly over node and edge representations regardless of the original feature sizes.

The processor consists of $L$ sequential message-passing layers, each of which simultaneously updates both edge and node latent representations. Within each processor layer, the edge update is performed first. For every edge $(i,j)$, the latent features of the source node $\mathbf{h_i}$, the destination node $\mathbf{h_j}$, and the current edge embedding $\mathbf{e_{ij}}$  are concatenated and passed through the edge encoder to produce an updated edge embedding. A residual connection is applied by adding the input edge embedding to the MLP output, yielding $\mathbf{e_{ij}'}=MLP_{edge} ([\mathbf{h_i}\|\mathbf{h_j}\|\mathbf{e_{ij}}])+\mathbf{e_{ij}}$. Following the edge update, messages are constructed for each edge by concatenating the neighboring node features $\mathbf{h_j}$ with the updated edge features $\mathbf{e_{ij}'}$. These messages are aggregated at each destination node using a summation operator, producing a fixed-size aggregated message vector for each node irrespective of its degree. The aggregated messages are then concatenated with the node's own latent features and passed through the node encoder. A residual connection is again applied, so the updated node representation is $\mathbf{h_i'}=MLP_{node}([\mathbf{h_i}\|\mathbf{m_i}])+\mathbf{h_i}$, where $\mathbf{m_i}$ denotes the aggregated incoming messages at node $i$. The residual connections in both the edge and node updates stabilize training and facilitate gradient flow through the processor layers. 

Finally, since the objective here is to predict nodal voltages, the SC-GNN formulation operates at the node level, with the ground node masked out since its voltage is a known constant of 0\,V. While graph-level predictions were also explored to directly obtain cable-level V-I characteristics, node-level predictions are more informative, as they provide spatially resolved voltage distributions across the circuit which are essential for protection-based system development where localized fault detection and response are critical. After $L$ rounds of message passing, the final latent node embeddings are passed through a decoder MLP consisting of three linear layers with ReLU activations between the first two and no activation on the final layer, allowing the output to span the full range of physically meaningful voltage values. The decoder produces a scalar output per node corresponding to the predicted nodal voltage.

The model is trained using a scale-normalized mean squared error (MSE) loss. Since nodal voltages in superconducting circuit simulations are typically sparse and concentrated near zero, raw MSE can vary significantly across batches, which leads to vanishing gradients and poor optimization. For the design space considered in this work, target values lie in the range $(0, 10^{-5})$, determined by the critical electric field $E_c = 1\,\mu$V/cm and the tape  length of 10\,cm.
Normalizing the MSE by the squared mean absolute target value therefore ensures stable and well-conditioned training. A mask is applied to exclude nodes whose voltages are prescribed as boundary conditions (ground node) from the loss computation. 
The loss is computed as $\mathcal{L_\text{data}}=\frac{1}{\mathcal{M}}\sum_{i\in \mathcal{M}}(\frac{\hat{v_i}-v_i}{s})^2$ , where $\mathcal{M}$ is the set of non-ground nodes, $\hat{v_i}$ and $v_i$ are the predicted and target voltages respectively, and $s=\frac{1}{\mathcal{M}} \sum_{i\in \mathcal{M}}\lvert v_i\rvert$ is the scaling factor equal to the mean absolute target voltage over the masked nodes.

\subsection{Physics-based loss}
To enforce physical consistency in the model predictions, a physics-informed loss term based on Kirchhoff's Current Law (KCL) is incorporated alongside the data-driven regression loss. This term penalizes violations of current conservation at each node. The current at each node is calculated using the predicted nodal voltages and the known constitutive relations of each circuit element. The voltage difference across each edge is computed from the predicted nodal voltages. Element-wise currents are then derived according to the governing relation for each element type, i.e., Ohm's law $I=V/R$  for resistive elements, and the prescribed source current for current source elements. 

Superconducting elements require fundamentally different treatment because their constitutive relation $V=V_c(I/I_c)^n$ cannot be directly inverted to obtain current from voltage in a numerically stable manner. For a resistor, the mapping from voltage to current is linear and bijective, yielding a unique and well-defined current for any voltage difference. For a superconductor operating in the sub-critical regime, however, the voltage across the element is vanishingly small because a large range of currents from zero to nearly $I_c$ all produce near-zero voltages when raised to a high power-law exponent $n$. Inverting the relation to obtain $I = I_c  (V/V_c )^{1/n}$ is therefore severely ill conditioned. Minute differences in the predicted voltage map to extremely different currents, and the gradient of the inverse with respect to voltage becomes vanishingly small, providing almost no useful learning signal to the network. To circumvent this, superconducting elements that are operating in the near-zero voltage regime are handled via a ``\textit{supernode merging}'' procedure. Specifically, if the predicted voltage difference across a superconducting edge falls below a threshold voltage, $\varepsilon_{\text{SC}}$ (set to $0.01\cdot V_c$), the two nodes connected by that edge are treated as electrically equivalent and merged into a single ``\textit{supernode}''. This reflects the physical reality that a superconductor carrying sub-critical current behaves as an ideal short circuit, imposing equal potential at both terminals. 

\begin{figure}
    \centering
    \includegraphics[width=1\linewidth]{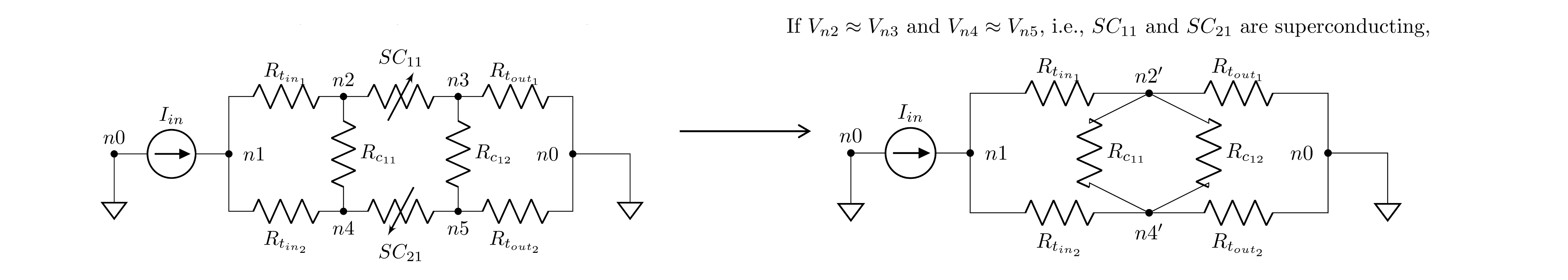}
    \caption{Supernode merging for KCL residual calculation. On the left is a schematic of a $2\times1$ tape stack with superconducting elements connecting nodes $n2$--$n3$ and $n4$--$n5$. When $I_{in}$ results in a current through the branches that is below the critical current of both $SC_{11}$ and $SC_{21}$, the voltage drop across each superconducting element is approximately zero, making $n2$ and $n3$ electrically equivalent, and similarly $n4$ and $n5$. The supernode merging exploits this equivalence by combining such node pairs, producing the reduced circuit shown on the right.}
\label{fig:kcl-mapping}
\end{figure}

Prior to the KCL evaluation, the edge list is filtered to unique directed edges with a fixed current flow convention,  such that the source node always has a smaller index than the target node, i.e. $u < v$ (except the edge between ground node and current input node). The supernode merging therefore becomes a direct node reassignment. First, a supernode mapping $\phi: i \mapsto i$ is initialized such that every node represents itself. For each short-circuited superconducting edge $(u, v)$, i.e. the $V_{u,v} < \varepsilon_{SC}$, the target node is redirected by setting $\phi(v) = u$, so that both nodes share the same representative. Once the map $\phi$ is fully constructed, every node appearing in the edge list is replaced by its supernode representative such that source node $u$ becomes $\phi(u)$ and target node $v$ becomes $\phi(v)$. This process is schematically descirbed in Figure.~\ref{fig:kcl-mapping}. KCL residuals are then accumulated at the supernode level rather than the individual node level. The ground node is excluded from the residual by construction, since its voltage is a fixed boundary condition. The KCL residual at each node (or supernode) $i$ is, therefore, defined as
\begin{equation}
 r_i = I_{\text{out},i} - I_{\text{in},i}    
\end{equation}

where the first part is the sum of currents leaving node $i$ and the second collects currents arriving. The residuals are normalised by the mean source current $I_{\text{avg}}$ present in the circuit to make the loss dimensionless and comparable across circuits of different operating scales. The physics-based loss is then the mean squared normalised residual over all non-ground nodes:

\begin{equation}
\mathcal{L}_{\text{physics}} = \frac{1}{\mathcal{M}} \sum_{i \in \mathcal{M}} \left( \frac{r_i}{I_{\text{avg}}} \right)^2    
\end{equation}

This term is added to the primary regression loss with a tunable weight $\lambda$, giving the combined training objective $\mathcal{L} = \mathcal{L}_{\text{data}} + \lambda \mathcal{L}_{\text{physics}}$. Since model predictions are largely random in the early epochs of training, the KCL residuals can be arbitrarily large and the physics loss can dominate the total loss, obscuring the regression signal. To mitigate this, the physics loss is only introduced after an initial training phase of fixed epochs, once the network has learned a basic fit to the data.

\subsection{Dataset generation}

Training a supervised surrogate model requires a dataset of input-output pairs generated by running the underlying simulation across a well-define representative range of circuit configurations and parameters. The training dataset is constructed from a mixed discrete--continuous design space. A full factorial grid is employed over the discrete variables $(n_t, m_t)$, while the remaining circuit parameters are sampled randomly. The parameter domain is summarized in Table~\ref{tab:doe}. Ranges for the parameters are selected from typical values observed in literature~\cite{martinez_electric-circuit_2020}. Additionally, for every $(n_t, m_t)$ pair, $K$ independent samples of the continuous parameters are drawn from uniform distributions over their respective ranges:
\begin{equation}
\begin{split}
&\forall (n_t,m_t)\in\{1,\dots,10\}\times\{1,\dots,6\}, \\
&\{R_t^{(k)}, R_c^{(k)}, I_c^{(k)}, n^{(k)}\}_{k=1}^K 
\overset{\text{i.i.d.}}{\sim} \\
&\quad \mathcal{U}(10^{-9},10^{-8}) \times
\mathcal{U}(10^{-9},10^{-8}) \times
\mathcal{U}(100,500) \times
\mathcal{U}(10,25)
\end{split}
\end{equation}

For each tape, a nominal critical current $I_c$ is first sampled and then perturbed to reflect realistic local variations along the tape and across the stack. Specifically, additive noise is introduced such that
\begin{equation}
I_c \leftarrow I_c + \epsilon,\quad \epsilon \sim \mathcal{N}(0, \sigma^2),
\end{equation}
where $\sigma = 3$\,A, corresponding to approximately 1\% of the mean source current $\mathbb{E}[I_c] = 300$\,A.

 To train an SC-GNN capable of generalizing across configurations ranging from $1\times1$ to $10\times6$, $K=20$ samples are drawn per configuration, yielding a training set of 2000 circuits. Validation uses $K=2$ samples per configuration, corresponding to 200 circuits in total. The sweep is discretized into $T$ operating points. At the $t$-th operating point ($t = 1,\dots,T$), the input current is defined as $I_{in} = \frac{t}{T} I_\text{target}$. For each circuit, this results in $T$ graph instances corresponding to different operating points along the current ramp. Across these graphs, the circuit topology remains unchanged. Only the edge features corresponding to the input current $I_{in}$ and the normalized operating ratio $I_{in}/I_c$ vary. For the current study, $T$ is set at 50. Once encoded as graphs, each circuit is stored as a Python dictionary containing node features, edge features, edge index, nodal voltages, and circuit metadata. For training, these are converted to PyTorch Geometric (PyG) format, where the numeric features are standardized. 

\begin{table}[]
\centering
\caption{Design space for circuit parameters.}
\label{tab:doe}
\renewcommand{\arraystretch}{1.2}
\begin{tabular}{l|l}
\hline
Variable & Range \\ \hline
$n_t$ & 1--10 \\ 
$m_t$ & 1--6 \\ 
$R_t\,(\Omega)$ & [$10^{-9} , 10^{-8}$] \\
$R_c\,(\Omega)$ & [$10^{-9} , 10^{-8}$] \\
$I_c\,(A)$ & [100 , 500] \\
$n$ & [10, 25] \\ \hline
\end{tabular}
\end{table}

\begin{figure}
    \centering
    \includegraphics[width=1\linewidth]{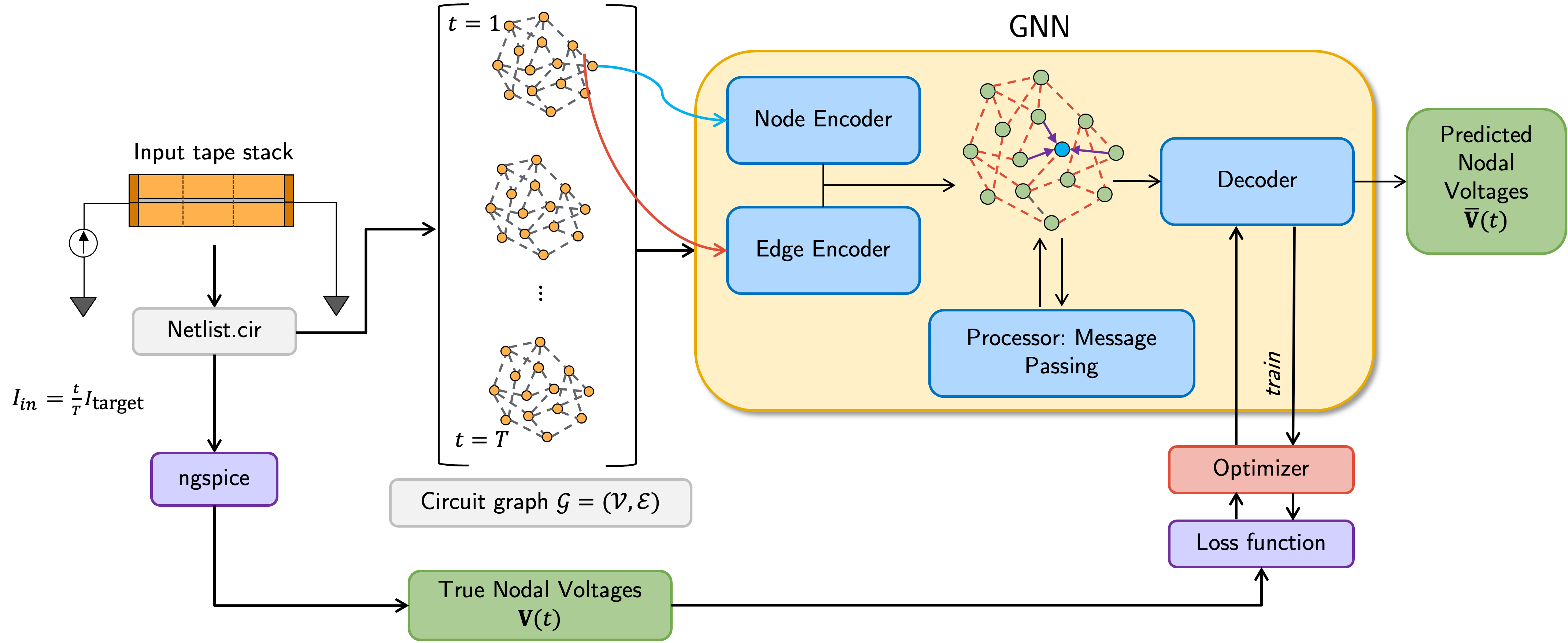}
    \caption{Overview of the SC-GNN framework. Starting from an input tape stack 
configuration, a SPICE netlist is generated and simulated using ngspice under 
a ramped input current $I_{in} = \frac{t}{T}I_{\text{target}}$ to produce true nodal 
voltages $\mathbf{V}(t)$. The circuit is simultaneously represented as a graph 
$\mathcal{G} = (\mathcal{V}, \mathcal{E})$ across $T$ current steps, which is 
passed through the \textit{encoder-processor-decoder} based GNN. The encoder comprising a node encoder and edge encoder produces latent embeddings that is sent through the message passing 
processor, and finally the decoder produces predicted nodal voltages $\bar{\mathbf{V}}(t)$. 
The model is trained by minimizing a loss function via gradient-based optimization.}
    \label{fig:overview}
\end{figure}

\subsection{Workflow and training}

An overview of the proposed SC-GNN framework for predicting nodal voltages is shown in Figure.~\ref{fig:overview}. The pipeline begins with a predefined architecture for a tape stack, specifying the number of tapes and segments per tape, from which a netlist file is generated. The netlist corresponds to SPICE formatted \textit{.cir} file for DC operating point simualtion where the current is swept to $I_{\text{target}}$ in $T$ steps. Simultaneously, the netlist is parsed into a circuit graph $\mathcal{G} = (\mathcal{V}, \mathcal{E})$, represented as a set of $T$ graph instances. The graph representation is passed to the SC-GNN consisting of the encoder, that projects raw node and edge features into latent embeddings, the processor implementing iterative message passing to propagate information across the graph topology, and finally the decoder that maps the updated node embeddings to the predicted nodal voltages. During training, predicted nodal voltages are compared against SPICE-simulated ground truth values via a loss function, and model parameters are updated through backpropagation using a gradient-based optimizer, defining the supervised training loop.

All model training was performed on the GPU partition of LBNL's Lawrencium cluster, utilizing a node equipped with an AMD EPYC 7742 processor (4 cores) and an NVIDIA A40 GPU. Hyperparameter optimization was performed using the Optuna framework~\cite{akiba_optuna_2019} to efficiently explore the model hyperparameter search space across two stages. In \textbf{Stage 1}, the GNN was optimized without a physics-based loss term, establishing the baseline model and representing the primary contribution of this work. \textbf{Stage 2}, which introduces a physics-informed loss to further constrain predictions, is currently ongoing and preliminary results are presented here to demonstrate the potential of incorporating physical priors into the learning objective. Further details on the hyperparameter search space and optimization procedure are provided in Appendix B.


\section{Results}

\subsection{Inference performance of baseline model}
\label{sec:inference_baseline}
\begin{table}[]
\caption{Optimized hyperparameters identified via Bayesian optimization for Stage 1 and 2.}
\centering
\label{tab:hyperparams}
\renewcommand{\arraystretch}{1.2}
\begin{tabular}{l|l|l}
\cline{1-3} 
Stage              & Parameter                       & {Value}    \\ \hline
\multirow{6}{*}{1} & Message passing layers ($L$)    & 6                            \\
                   & Hidden dimension ($d_h$)        & 64                           \\
                   & Learning rate                   & 6.89$\times 10^{-4}$                     \\
                   & Weight decay                    & 2.85e$\times 10^{-6}$                     \\
                   & Batch size                      & 64                           \\
                   & Optimizer                       & Adam                         \\ \hline
\multirow{2}{*}{2} & Warmup epochs for physics loss  & {50}       \\
                   & Physics loss weight ($\lambda$) & {5.62e$\times 10^{-3}$} \\ \hline
\end{tabular}
\end{table}

\begin{figure}
    \centering
    \includegraphics[width=1\linewidth]{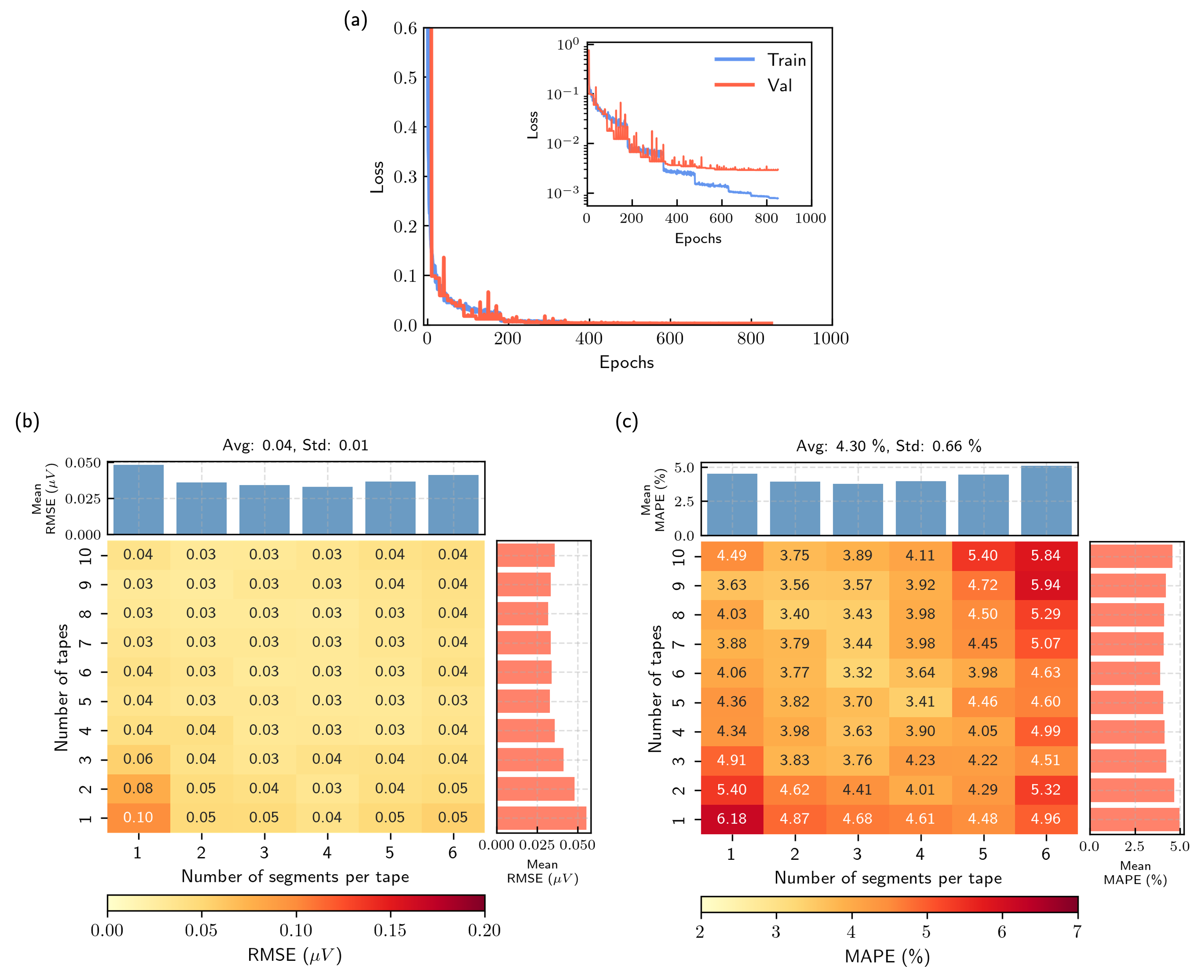}
    \caption{Baseline model performance. (a) Training and validation loss curves showing the evolution of the baseline SC-GNN over 850 epochs with the inset showing the same on logarithmic scale. Inference performance evaluated on the 
held-out test set across the full topology space ($1\times1$ to $10\times6$ 
tapes $\times$ segments per tape). (b) RMSE ($\mu$V) and (c) MAPE (\%) heatmaps, 
where each cell represents the mean error averaged over $K=50$ independently 
sampled circuit parameter sets. Bar charts alongside each axis show the row-wise, i.e., tape-wise,
(red) and column-wise, i.e., segment-wise, (blue) marginal distributions of the respective error metric.}
    \label{fig:gnn_result}
\end{figure}

The best Stage 1 model, i.e., baseline model, trained using the optimized hyperparameters in 
Table~\ref{tab:hyperparams}, converged after 850 epochs with early stopping 
(patience of 10 epochs on validation loss) and a total wall time of 3.58\,h. Figure.~\ref{fig:gnn_result}(a) shows the loss curves with the training and validation loss plotted over epochs. The validation loss closely follows the training loss indicating the model generalizes well and there is no significant overfitting. 
To evaluate the performance of the trained SuperCond-GNN surrogate, inference 
results are examined on the held-out test data spanning the full topology space 
of tapes and segments per tape, with circuit parameters sampled uniformly within 
the design space using $K=50$ samples per topology, yielding 3{,}000 circuits in 
total. Model predictions were evaluated using the root mean squared error 
(RMSE $= \sqrt{{\sum_{i\in\mathcal{M}}(v_i-\hat{v}_i)^2}/{\mathcal{M}}}$) and 
the mean absolute percentage error 
(MAPE $= \frac{1}{\mathcal{M}}\sum_{i\in\mathcal{M}}\left|\frac{v_i - \hat{v}_i}{v_i}\right| \times 100$). The heatmaps in 
Figure~\ref{fig:gnn_result}(b) and (c) show the RMSE and MAPE, respectively, 
of the SC-GNN evaluated across the full topology space of $1\times1$ to 
$10\times6$. Each cell represents the mean error averaged over the 50 
independently sampled test sets, providing a robust estimate of model performance 
across circuit configurations.

Overall, the model achieves a mean RMSE of $0.04\,\mu$V ($\pm\,0.01\,\mu$V) and 
a mean MAPE of $4.30$\% ($\pm\,0.66$\%), demonstrating strong predictive accuracy 
across the majority of circuit configurations. Performance is broadly consistent 
across mid-to-large topologies, with RMSE values typically in the range 
$0.03$--$0.04\,\mu$V and MAPE values between $3$--$5$\%, indicating that the 
SC-GNN generalizes well to unseen parameter combinations within these 
configurations. Notably, performance does not degrade monotonically with circuit size, with 
several larger topologies achieving comparable or lower error than mid-range 
configurations, suggesting the message passing framework scales effectively with 
graph size. The marginal distributions shown in the bar charts alongside each axis 
summarize performance trends across individual dimensions. The row-wise marginal 
(red) captures performance as a function of the number of tapes, while the 
column-wise marginal (blue) captures trends across the number of segments per 
tape. Along the tape axis, mean MAPE remains broadly stable for 
configurations with 4 or more tapes. Along the segment axis, the marginal 
distributions are comparatively flat with no strong systematic trend, though modest 
increases in MAPE are observed at higher segment counts for lowest and highest tape configurations. 

A modest degradation in performance is observed for circuits with low tape counts, 
where RMSE reaches up to $0.10\,\mu$V and MAPE up to $6.18$\% for the 
$1\times1$ topology. Importantly, this degradation is driven primarily by tape 
count rather than segment count, as the 1-tape row exhibits consistently elevated 
errors across all segment configurations (MAPE ranging from $4.48$--$6.18$\%), 
rather than being confined to small topologies in both dimensions simultaneously. 
This may reflect the greater sensitivity of small circuits to parameter variations, 
where a single segment transitioning between the superconducting and resistive 
state has a proportionally larger effect on the overall operating point. 
Additionally, the graph structure for these small circuits provides fewer 
neighboring nodes from which to aggregate information during message passing. This 
limited local context may reduce the expressiveness of the learned node 
representations, making accurate prediction harder despite the apparent simplicity 
of these configurations. Regardless, the performance on the single tape topologies are of less concern since they do not qualify as a stack and can therefore be avoided in future training campaigns. A similar, though less pronounced, degradation is observed at high tape counts, with MAPE values reaching up to $5.94$\% for the $9\times6$ topology, particularly at larger segment counts, while RMSE remains broadly consistent. This discrepancy is partly attributable to the metric computations itself where larger circuits contain more nodal points, increasing the likelihood of 
encountering nodes with small absolute voltage values where even small absolute 
errors yield disproportionately large percentage contributions to MAPE, while 
the RMSE remains relatively unaffected as it aggregates absolute errors uniformly 
across all nodes. Regardless, the higher MAPE is also a result of the increased complexity of larger graphs, where the fixed number of message passing layers $L=6$ limits the receptive field of each node, potentially preventing full aggregation of 
information across the graph diameter. 

\begin{figure}
    \centering
    \includegraphics[width=1\linewidth]{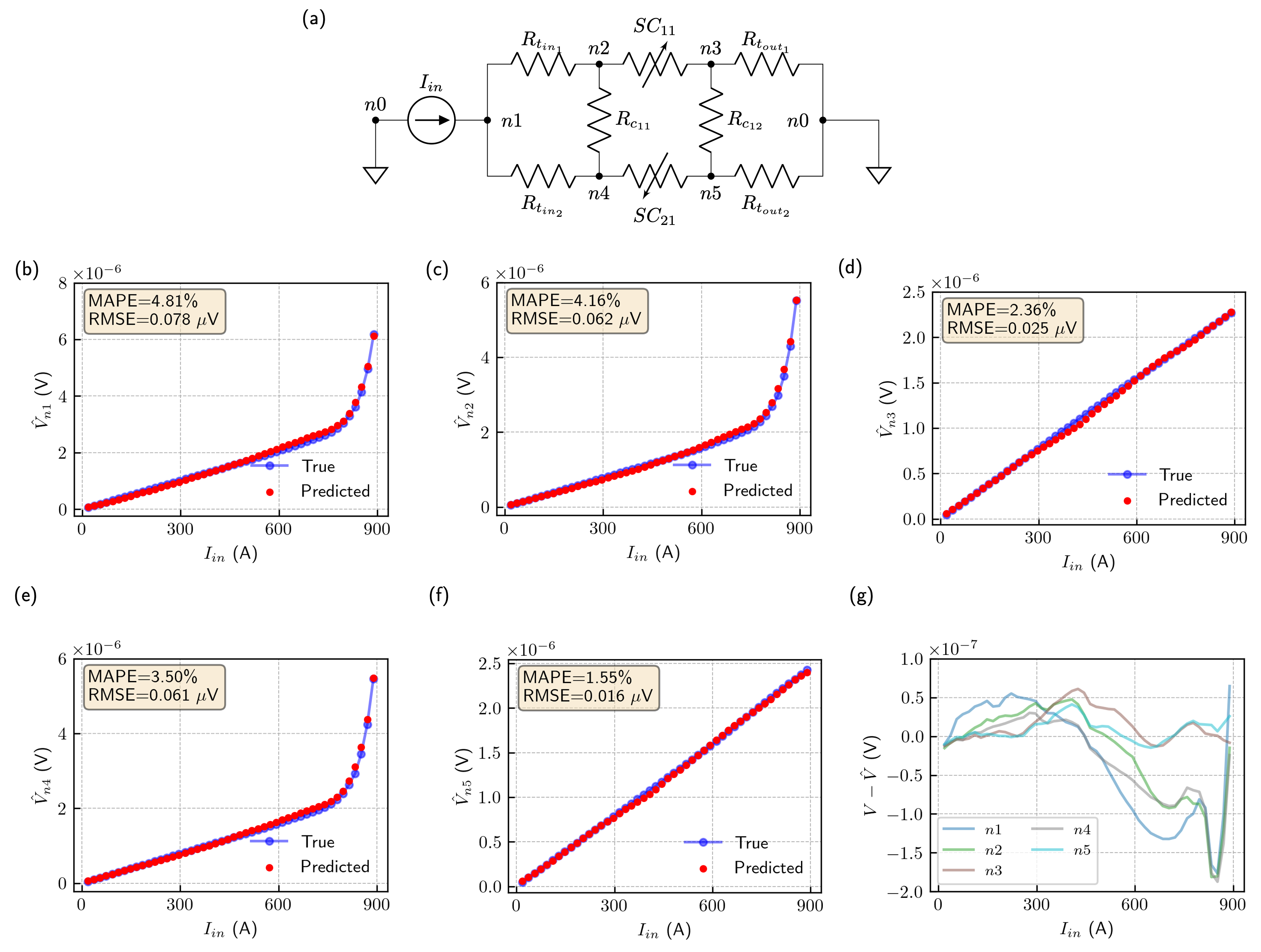}
        \caption{SC-GNN inference results for a representative $2\times2$ tape stack. 
(a) Equivalent circuit schematic showing the nodal structure with 2 tapes and 
2 segments per tape, with nodes $n1$--$n5$ labelled ($n0$ is the ground). (b)--(f) Predicted versus 
true nodal voltages $\hat{V}_{n1}$ to $\hat{V}_{n5}$ as a function of input 
current $I_{in}$, with RMSE and MAPE reported for each node. (g) 
Residual errors $V - \hat{V}$ across all nodes as a function of $I_{in}$, 
illustrating the distribution and magnitude of prediction errors. The nonlinear 
voltage response at high current values, associated with the superconducting to resistive transition, is well captured by the model across all nodes.}
    \label{fig:gnn_result_sample}
\end{figure}

Figure.~\ref{fig:gnn_result_sample} shows the predicted and true nodal voltages for a representative 2-tape, 2-segment circuit (panel a) across a sweep of cable current values. For each node, the SC-GNN predictions (red) closely track the true ngspice solutions (blue) across the full current range, including the nonlinear transition region between approximately $600-900$\,A where segments begin switching from the superconducting to the resistive state. MAPE values across the five nodes range from 1.55\% to 4.81\%, and corresponding RMSE values between 0.016 and 0.078\,$\mu V$, with the highest errors at the input node. Additionally, the input node voltages correspond to the voltage response of the tape stack as a whole. An RMSE of 0.078\,$\mu$V translates to an equivalent electric-field error of 0.0078\,$\mu$V/cm for a 10\,cm tape. This is less than 1\% of the conventional 1\,$\mu$V/cm critical-current criterion, indicating that the surrogate error is negligible relative to experimentally relevant voltage levels. Panel (g) shows the residuals across all nodes as a function of cable current, revealing that prediction errors are largest within the transition region, where the voltage response is most nonlinear. Despite this, the residuals remain small in magnitude, demonstrating that the SC-GNN captures the essential physics of the DC operating point across both the superconducting and resistive regimes.


\subsection{Computational performance and scalability}

\begin{figure}
    \centering
    \includegraphics[width=\linewidth]{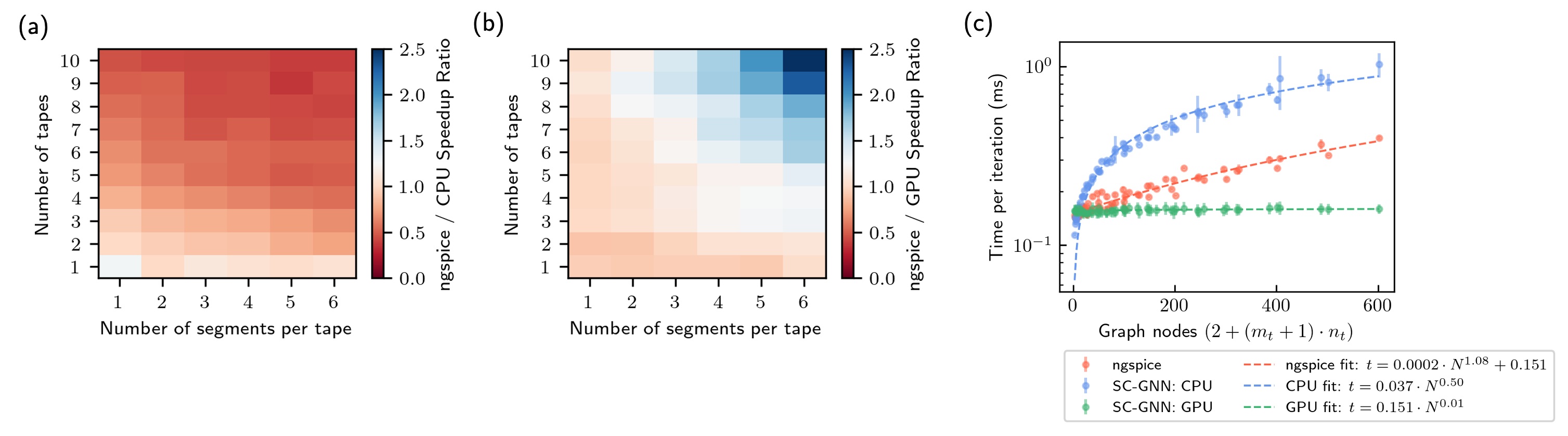}
    \caption{Computational performance of the SC-GNN surrogate relative to ngspice 
across the full topology space. (a) ngspice-to-CPU speedup ratio and (b) 
ngspice-to-GPU speedup ratio heatmaps across the topology space of $1\times1$ 
to $10\times6$ tapes $\times$ segments per tape, where values greater than 1 
indicate the SC-GNN is faster than ngspice. (c) Time per iteration (ms) as a 
function of graph size ($2 + (m_t+1)\cdot n_t$ nodes) for ngspice, SC-GNN on 
CPU, and SC-GNN on GPU, with power-law fits indicating the scaling behaviour 
of each method. All benchmarks were performed on an AMD EPYC 7742 CPU with an 
NVIDIA A40 GPU and averaged over three runs.}
    \label{fig:cost}
\end{figure}

Benchmarking study on a hold-out test dataset was carried out across CPU inference, GPU inference, and ngspice. All benchmarks were performed on the same hardware node (AMD EPYC 7742, 4 cores, 
NVIDIA A40 GPU) to ensure a fair comparison. Each benchmark was repeated three 
times and the reported values represent the mean over these runs. ngspice timing was measured as the total wall-clock time of the DC sweep divided by the number of operating points, with file I/O and netlist generation excluded, capturing only the time taken by the ngspice solver itself. SC-GNN inference timing includes data preparation, transfer of node and edge features to the target device, batching of all operating points for a given circuit into a single forward pass, and model inference. For the inference, warmup passes were performed prior to timing for both CPU and GPU to exclude one-time compilation and JIT overhead from the reported measurements. It is important to note that ngspice does not natively support complete multi-core parallelism or GPU acceleration (limited to certain rudimentary elements such as capacitors and resistors), and all ngspice runs were therefore executed on a single CPU core. 

Figure.~\ref{fig:cost} presents a comprehensive benchmarking comparison between SC-GNN surrogate inference on CPU and GPU, and ngspice, across the full topology space. The heatmaps in panels (a) and (b) show the pairwise speedup ratios across the topology space. CPU inference times range from approximately 0.1\,ms/iter for small circuits to around 1.0\,ms/iter for the largest topologies, growing monotonically with circuit size. GPU inference is substantially faster and remains largely flat across the topology space at approximately 0.15\,ms/iter. ngspice times range from approximately 0.2\,ms/iter for small circuits to 0.14--0.4\,ms/iter for the largest topologies, following a similar trend to CPU inference. Panel (a) shows that CPU inference offers no consistent speedup over ngspice across the circuit sizes tested, with the ngspice/CPU ratio remaining close to or below 1 for most configurations. However, as seen in panel (b),GPU inference is competitive with or faster than ngspice across the majority of configurations, achieving a peak ngspice/GPU speedup of 2.5$\times$ for larger circuits. This is consistent with the GPU's ability to exploit parallelism across the batched operating points more effectively as graph size increases. 

Notably, GPU inference time remains largely insensitive to circuit size. This is clearer in panel (c) which shows inference time per operating point as a function of graph size (total nodes) on a logarithmic scale, with power law fits revealing distinct scaling behaviours across the three approaches. GPU inference exhibits near-constant scaling with an exponent of 0.008, remaining at approximately 0.15\,ms/iter across the entire range of circuit sizes tested. ngspice scales linearly with an exponent of 1.08, consistent with the direct matrix solving characteristic of SPICE simulators, while CPU inference scales most steeply with an exponent of 0.496. These scaling exponents directly imply that the GPU surrogate will become increasingly advantageous over both ngspice and CPU inference as circuit complexity grows to larger magnet configurations. While this work represents a proof-of-concept on circuit representations of tape stacks of up to 10 tapes and 6 segments per tape, larger and more complex configurations are expected to yield more pronounced speedup advantages on GPU as indicated by the empirical scaling analysis.

\begin{figure}
    \centering
    \includegraphics[width=\linewidth]{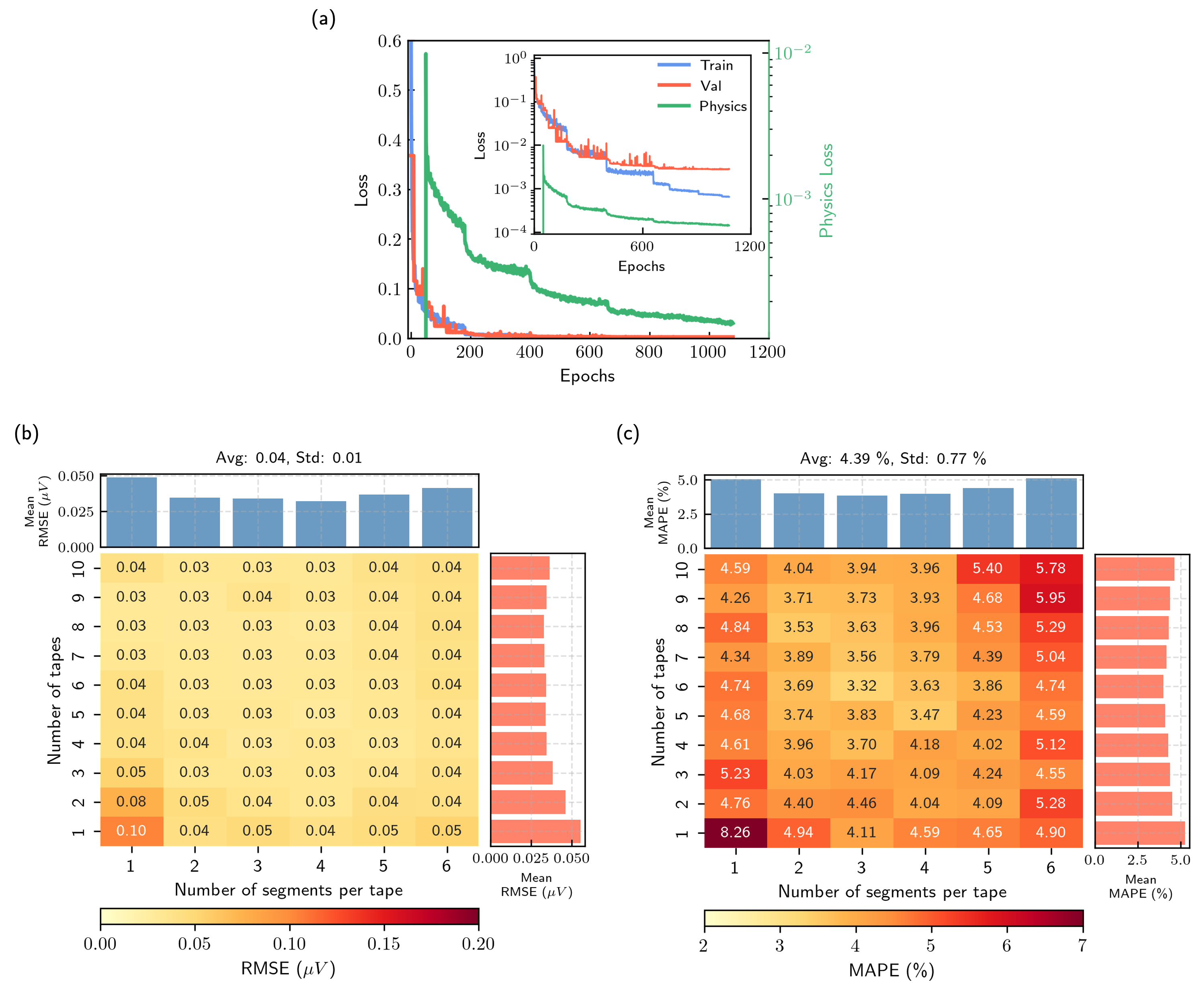}
    \caption{SC-GNN inference results following physics-informed Stage 2 training. 
(a) Evolution of the training loss, validation loss, and physics loss 
($\mathcal{L}_\text{physics}$) over 1080 epochs, with the inset showing the 
convergence behaviour on a logarithmic scale. (b)--(c) Inference performance 
on the held-out test set across the full topology space ($1\times1$ to 
$10\times6$ tapes $\times$ segments per tape): (b) RMSE ($\mu$V) and (c) MAPE 
(\%) heatmaps, where each cell represents the mean error averaged over $K=50$ 
independently sampled circuit parameter sets. Bar charts alongside each axis 
show the row-wise (red) and column-wise (blue) marginal distributions of the 
respective error metric. The physics-regularized model achieves a mean RMSE of 
$0.04\pm0.01\,\mu$V and a mean MAPE of $4.39\pm0.77$\%, comparable to the 
baseline SC-GNN.}
    \label{fig:gnn_w_phy_result}
\end{figure}

\subsection{Evaluation of physics-based regularization} 

Figure.~\ref{fig:gnn_w_phy_result} (a) shows the training curves along with the unweighted physics loss evolution for the SC-GNN with physics-based loss using parameters listed in Table.~\ref{tab:hyperparams}. The physics-informed Stage 2 training yields a modest but measurable improvement 
in best validation loss from $2.87\times10^{-3}$ to $2.75\times10^{-3}$, 
corresponding to a $\sim$4\% reduction in normalized error.  However, this comes 
at the cost of nearly doubled training time from 3.58\,h to $\sim$7\,h. 
Subsequent inference on the held-out test set (same as Section~\ref{sec:inference_baseline}) 
reveals that the physics-regularized model performs comparably to the baseline 
SC-GNN, with both variants achieving a mean RMSE of $0.04\pm0.01\,\mu$V and a 
mean MAPE of $4.39\pm0.67$\%, as seen in Figure.~\ref{fig:gnn_w_phy_result}(b) and (c). Rather than indicating a limitation of the 
physics-informed approach, this suggests that the SC-GNN architecture already 
possesses a strong inductive bias toward the underlying electrical physics through 
its graph topology alone, such that explicit KCL regularization in its current implementation provides limited 
additional benefit on the test set, beyond what the supervised loss achieves.

While the physics-informed approach provides theoretical robustness, the current formulation introduces a significant computational overhead compared to a similar architecture without physics loss, despite optimal tuning of the hyperparameters. This computational bottleneck is primarily attributed to the circuit remapping to supernodes, which is currently necessary to calculate the implicit KCL residuals from the predicted nodal voltages. It is worth noting that KCL was deliberately selected over Kirchhoff's Voltage Law (KVL) for this regularization, since the message aggregation inherent in the SC-GNN's message passing layers effectively mirrors KVL satisfaction. Specifically, in a circuit graph where edge features encode branch voltages, the aggregation of messages along any closed loop naturally enforces voltage consistency, as the sum of voltage drops around any cycle is implicitly constrained to zero through the iterative message passing updates. An explicit KVL loss would therefore provide redundant information with limited incremental value to the training process.

Given the computational cost of the supernode mapping and the strong performance 
of the unregularized baseline, alternative strategies for integrating physics 
constraints are being explored. A primary future direction is reformulating the 
framework as an edge-based graph problem in which branch currents are directly 
predicted, allowing KCL residuals to be computed directly without the supernode 
remapping step. Another alternative is to dynamically weight the physics loss 
to target nodes with the highest KCL violations, rather than averaging uniformly 
across the circuit. This may provide more targeted gradients and better realize the 
potential of physics-informed learning in this domain.

\subsection{Evaluating model generalizability}

\begin{figure}
    \centering
    \includegraphics[width=\linewidth]{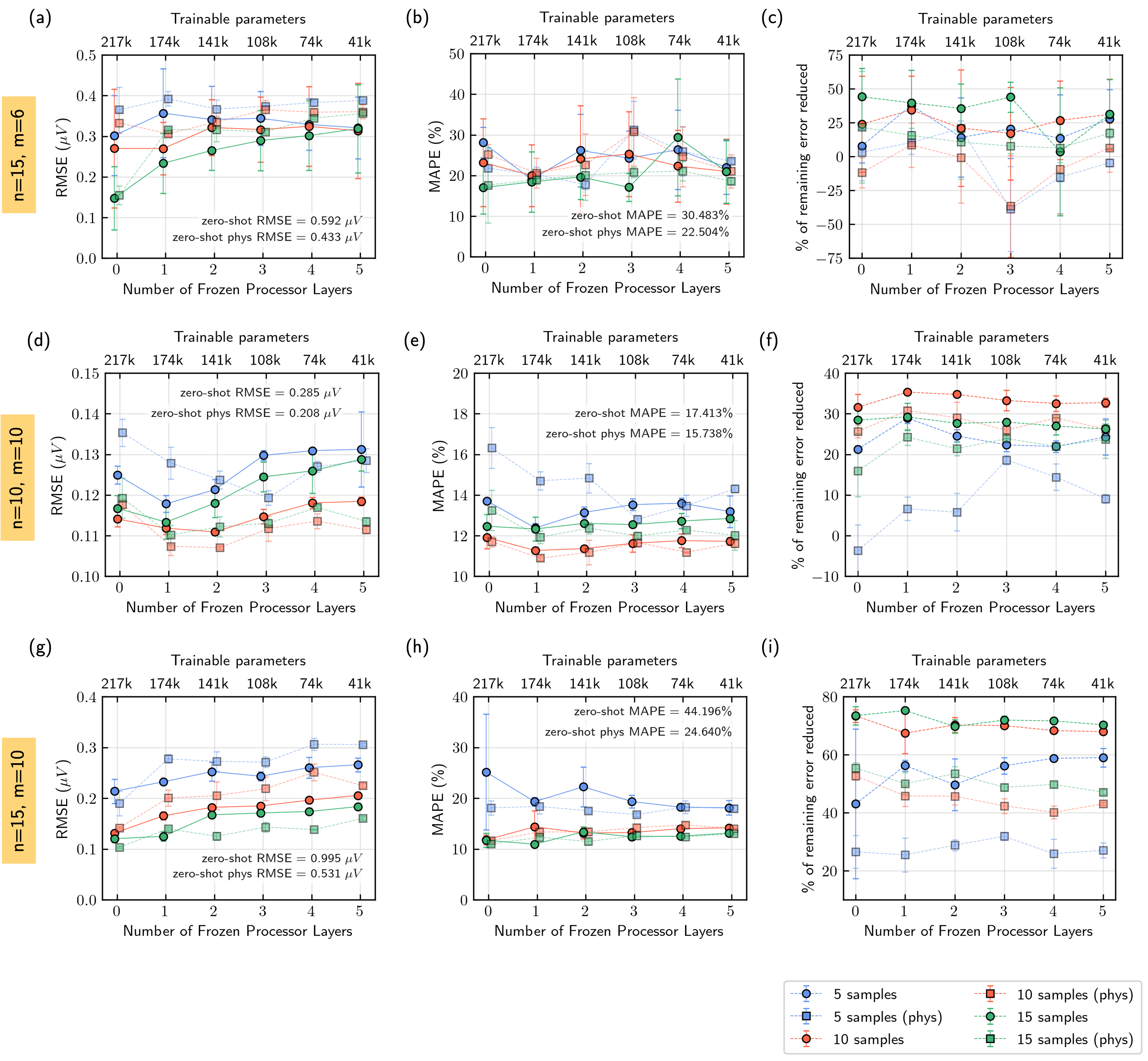}
    \caption{Few-shot generalization results for three out-of-distribution (OOD) 
tape stack topologies: $15\times6$ (a--c), $10\times10$ (d--f), and $15\times10$ 
(g--i). For each topology, RMSE ($\mu$V) (a, d, g), MAPE (\%) (b, e, h), and 
percentage of remaining error reduced relative to zero-shot performance (c, f, i) 
are shown as a function of the number of frozen processor layers, with the number 
of trainable parameters indicated along the top axis. Results are shown for 5, 10, 
and 15 fine-tuning samples, with and without physics-informed loss. Horizontal annotations indicate 
zero-shot RMSE and MAPE for both standard and physics-informed inference. All 
results are averaged over three runs on 10 test circuits.}
    \label{fig:few_shot}
\end{figure}

A key requirement for the broader applicability of SC-GNN to unseen magnet 
configurations is its ability to generalize. To assess this, generalizability is 
evaluated on larger tape stack topologies representing out-of-distribution (OOD) 
conditions. Zero-shot generalization is first tested on these unseen topologies, 
followed by few-shot fine-tuning. For the latter, training samples are 
progressively increased and the pre-trained surrogate is fine-tuned under six 
configurations: all processor layers unfrozen, followed by iterative freezing of 
processor layers from 1 to 5, the latter leaving only the final layer 
and decoder trainable. Freezing a layer means its weights are held fixed during training, preventing them from being updated. Both standard and physics-informed fine-tuning variants 
are evaluated. Three OOD tape stack topologies are considered: (1) $15\times6$, 
(2) $10\times10$, and (3) $15\times10$. The pretrained model was trained on tape 
stack configurations from $1\times1$ to $10\times6$. Case (1) corresponds to an 
extension in the number of tapes, introducing more parallel branches in the 
circuit. Case (2) represents a horizontal expansion in which the number of 
segments per branch is increased while the number of parallel branches remains 
within distribution. Case (3) represents the most challenging scenario, as the 
tape stack configuration exceeds the training distribution in both dimensions 
simultaneously. In terms of graph size, the pretrained SC-GNN was exposed 
to graphs with 3 to 602 nodes during training and is now applied to graphs of 
1352, 1002, and 2252 nodes for cases (1), (2), and (3), respectively. For each 
case, reported metrics are averaged over three runs on 10 test circuits.

\subsubsection{Zero-shot inference} For zero-shot inference, the pretrained model is 
applied directly without any fine-tuning. Average MAPE values for $15\times6$, 
$10\times10$, and $15\times10$ are 30.49\%, 17.41\%, and 44.20\%, respectively, while 
corresponding average RMSE values are 0.59\,$\mu$V, 0.28\,$\mu$V, and 0.99\,$\mu$V. The 
physics-informed zero-shot variant achieves lower errors in all three cases, 
with RMSE of $0.43\,\mu$V, $0.21\,\mu$V, and $0.53\,\mu$V and MAPE of 
$22.50$\%, $15.74$\%, and $24.64$\% for cases (1), (2), and (3), 
respectively, suggesting that KCL regularization improves generalization to 
OOD topologies even without fine-tuning. Zero-shot performance is considerably 
worse than that reported in Section~\ref{sec:inference_baseline} for 
in-distribution test data, as expected. Among the three OOD cases, zero-shot 
performance is best for $10\times10$, where only the number of segments per tape 
is increased. This is expected for two reasons. First, edge features remain 
within distribution since none are dependent on the number of segments per 
branch. Second, this case corresponds to a comparatively moderate increase in 
graph size without affecting the tape stack behavior much, making it the simplest of the three scenarios. The variance across 
iterations is also considerably tighter for this case. Performance on $15\times6$ 
ranks second, as increasing the number of tapes alters the physical configuration 
of the tape stack by increasing current-carrying capacity and introducing greater 
current sharing. These two sources of OOD shift compound in complexity when 
combined, which explains why zero-shot performance is worst for $15\times10$.

\subsubsection{Few-shot fine-tuning} 
Fine-tuning generally improves performance over zero-shot inference across the 
three OOD topologies, as reflected in the percentage of remaining error reduced 
shown in Figure~\ref{fig:few_shot}(c, f, i), though the degree of improvement 
varies considerably across cases and fine-tuning configurations.  These improvements are further increased
when the more samples are provided for training which also reduces the observed variance across the runs. For $15\times6$ and $10\times10$, the benefit of increasing from 
10 to 15 samples is less clear, with some configurations exhibiting an opposite 
trend, suggesting sensitivity to the specific parameter combinations sampled in 
the small training sets. However, for $15\times10$, 
increasing the number of training samples consistently yields further improvement, 
with 15 samples outperforming 5 and 10 across most configurations and variance reducing 
with more data. Nevertheless, the results demonstrate that SuperCond-GNN 
can adapt to unseen topologies through fine-tuning on relatively small datasets.

Regarding the effect of processor layer freezing, no single strategy consistently 
dominates across all cases and firm conclusions cannot be drawn at this stage. As the number of frozen layers increases, the number of trainable parameters is approximately halved with each step, yet performance does not degrade monotonically.
For $15\times6$, variance across freezing configurations is particularly high, 
with some configurations performing worse than zero-shot inference, indicating 
strong sensitivity to which layers are left trainable for this topology. For 
$10\times10$, error reduction is consistently positive and variance is tighter 
across freezing levels, suggesting that adaptation to horizontal segment expansion 
is a more tractable fine-tuning task. The error reduction is also 
consistently positive across configurations for $15\times10$. More generally, minimal freezing appears favorable across cases, preserving the model's ability to adapt while still benefiting from the pretrained representations.  However, given the high variance and limited number of 
fine-tuning samples, the effect of freezing configuration remains inconclusive 
and warrants more systematic investigation.

An important distinction especially emerges between the zero-shot and fine-tuning behaviour 
of the physics-informed variant. While the physics-informed model achieves 
consistently lower zero-shot errors across all three OOD cases, the gains from 
fine-tuning are comparatively smaller than those observed for the baseline. For 
$15\times6$ specifically, fine-tuning the physics-informed model degrades 
performance relative to its own zero-shot baseline, in contrast to the baseline 
model which fine-tunes readily despite its higher zero-shot error. This suggests 
a trade-off between zero-shot generalization and fine-tuning adaptability, where 
KCL regularization improves out-of-distribution robustness but may constrain 
plasticity during adaptation. Notably, this degradation is not observed for 
$15\times10$ despite both cases exceeding the training distribution in tape count, 
which remains an open question. Generalizability 
to larger circuits is expected to benefit from increased model depth, as additional 
processor layers would better capture the long-range dependencies present in 
larger graphs, and from more diverse training topologies to improve the robustness 
of few-shot adaptation.

\section{Conclusion}

This work presented SuperCond-GNN, a graph neural network-based surrogate model designed as a system-agnostic framework for predicting nodal voltage distributions in HTS magnets. As a proof of concept, tape stack circuits were represented as equivalent electrical networks and converted into graph representations, from which nodal voltages were learned using a message passing GNN trained on ngspice-generated data across the full topology space of $1\times1$ to $10\times6$ configurations, where $n_t \times m_t$ denotes the number of tapes and segments per tape respectively.

The proposed surrogate achieved a mean RMSE of $0.04\,\mu$V and a mean MAPE of 
$4.30$\% on the held-out test set, demonstrating strong predictive accuracy across 
the full topology space. Performance degraded only modestly at the lowest tape 
counts and at the upper extreme of the design space, consistent with the limited 
local graph context available to the message passing layers in these regimes. Crucially, these error margins translate to an equivalent electric-field error that is well below 1\% of the conventional $1\,\mu$V/cm critical-current criterion, demonstrating that the model's accuracy is within the tolerances required for practical experimental measurements. SuperCond-GNN also exhibits near-constant scaling with graph size
$\propto N^{0.008}$, compared to the linear scaling of ngspice $\propto N^{1.08}$, where $N$ is the number of graph nodes. This demonstrates 
the scalability advantage of the surrogate for larger circuits and establishes it 
as a viable tool for downstream applications including design space exploration, 
current sharing analysis, optimization, and real-time simulation.

The incorporation of a physics-informed loss targeting KCL as a regularizer 
yielded a modest improvement in validation loss but comparable test set 
performance to the baseline, suggesting that the graph topology alone provides 
a strong inductive bias toward the underlying electrical physics. However, the 
physics-informed variant demonstrated a more pronounced advantage in zero-shot 
generalization to OOD tape stack topologies, consistently achieving lower RMSE 
and MAPE than the baseline across all three OOD cases evaluated. Both variants 
were further assessed under few-shot fine-tuning across varying dataset sizes and 
processor layer freezing configurations, demonstrating that SuperCond-GNN can 
adapt to unseen topologies with as few as 5 to 10 training samples. 

Future work will scale SuperCond-GNN to larger tape stacks and more complex magnet topologies. To support this scaling, the framework will be enhanced with deeper, attention-driven architectures and more efficient physics-informed training, such as dynamic loss weighting for KCL violations. Ultimately, the long-term objective is to transition from simulation to real-world magnet state estimation. By incorporating physical coil measurements, the model can be deployed onto hardware platforms like FPGAs to serve as a real-time voltage predictor for practical magnet protection.

\section*{Appendices}
\begin{appendices}
\section{Ablation studies}

This section reports the ablation studies performed to justify key architectural design choices for the graph representation of tape stack circuits. Specifically, the following features are evaluated:
\begin{enumerate}
\item One-hot encoding of circuit elements in the edge features (as opposed to weighted encoding).
\item Normalized distance from the current source as a node feature.
\item Normalized distance from the ground as a node feature.
\item Node degree as a node feature.
\end{enumerate}

To isolate the effects of each feature, changes were applied to the baseline model one at a time. The combinatorial influence of altering multiple features simultaneously was not studied, as their interactions were assumed to be minimal. The study utilized a smaller design space comprising of $1\times1$ to $4\times4$ tape stack configurations, maintaining an 80:10:10 ratio for the training, validation, and test splits. Each model configuration was trained three times with independent initializations, and results are reported as averages across these runs. The baseline model (M0) utilizes one-hot encoding for edges and includes all aforementioned node features. To test the effect of edge encodings, M0 was compared against two weighted variants: M1 utilizes hierarchical edge weights (resistor: 1, current source: 2, superconductor: 3), and M2 utilizes extreme edge weights (resistor: 1, current source: 5, superconductor: 10). To evaluate the importance of node features, features were systematically removed from the baseline. M3 removes the distance from the current source, M4 removes the distance from the ground, and M5 removes the node degree.

\begin{figure}
    \centering
    \includegraphics[width=\linewidth]{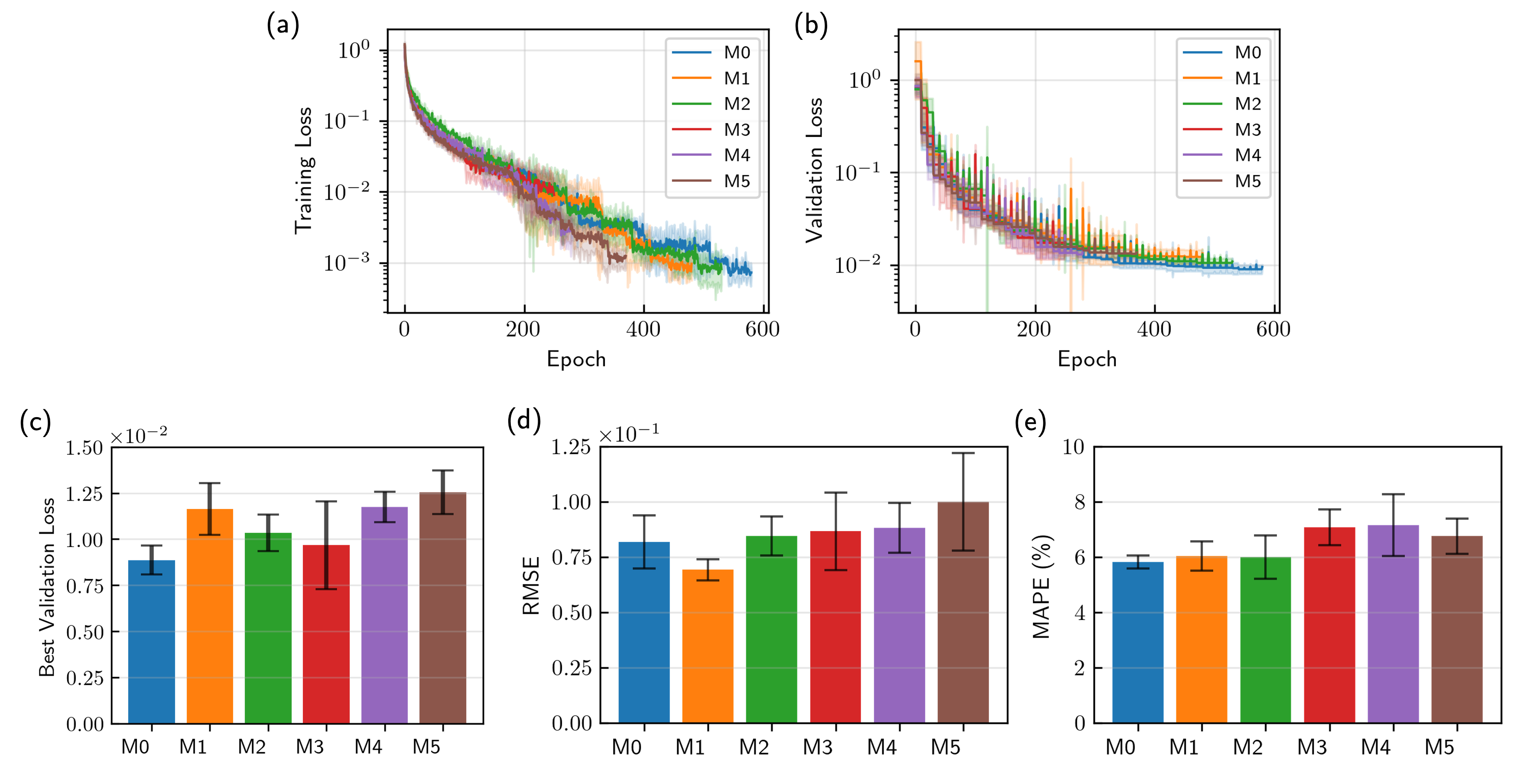}
    \caption{Ablation study results on $1\times1$ to $4\times4$ tape stack configurations across model variants M0--M5 (M0: Baseline, M1: Hierarchical edge feature, M2: Extreme edge features, M3: Normalized distance from current source removed from node feature, M4: Normalized distance from ground removed from node feature, and M5: Node degree removed from node features) . (a) Training loss and (b) validation loss curves over for all 
variants, shown on a logarithmic scale with shaded regions indicating the standard 
deviation across runs. (c) Best validation loss, (d) test RMSE ($\mu$V), and (e) 
test MAPE (\%) for each variant, with error bars denoting the standard deviation. 
All variants converge to comparable training and validation loss trajectories, 
with differences in generalization performance more apparent in the test metrics.}
    \label{fig:ablation}
\end{figure}

Figures.~\ref{fig:ablation}(a) and (b) illustrate the training and validation loss curves across all model variants. M0 exhibits the most stable loss reduction and trains for the longest duration, with M2 performing as a close second. Notably, all configurations exhibit a similar degree of overfitting, with training loss converging to approximately one order of magnitude below the validation loss.

Panel (c) presents the best validation loss achieved by each variant, with error bars reflecting variance across three training runs. M0 yields the strongest performance overall. Modifications to the edge representation consistently degrade performance. The hierarchical edge weight encoding (M1) increases validation loss by 31.3\% relative to the baseline, while the extreme edge weight encoding (M2) results in a 16.6\% increase. Removing node features similarly impairs performance. Omitting the normalized distances from the current source (M3) and ground (M4) leads to validation loss increases of 9.1\% and 32.5\%, respectively. Together, these features encode the primary determiners of current flow through a node. The most severe degradation arises from removing node degree (M5), a feature that captures local topological context, resulting in a 41.6\% increase in validation loss.

Test performance, evaluated on held-out data, is summarized in panels (d) and (e). Interestingly, M1 achieves the lowest RMSE despite its comparatively higher validation loss, with M0 remaining a close second. This apparent contradiction may be explained by the scale of the target values, which span several orders of magnitude. RMSE penalizes absolute errors and is therefore dominated by the largest predictions in this range, meaning a model that better fits high-magnitude samples can achieve lower RMSE while performing worse overall. MAPE, by contrast, evaluates relative error uniformly across the prediction range and is consequently a more informative metric in this setting. By this measure, M0 achieves the best performance. The edge encoding variants M1 and M2 attain comparable MAPE values to M0, albeit with slightly higher variance, suggesting that the choice of edge encoding may have a marginal and test-case-dependent effect on final accuracy. In contrast, the performance degradation caused by removing node features remains pronounced in testing, reinforcing their importance to the model's predictive capacity.

\section{Hyperparameter optimization}

Hyperparameter optimization was performed using the Optuna framework~\cite{akiba_optuna_2019}, employing a Tree-structured Parzen Estimator-based Bayesian optimization to efficiently explore a nine-dimensional hyperparameter search space. Optimization was conducted in two sequential stages. Stage 1 focused on standard architectural and training hyperparameters, comprising: 
\begin{itemize}
    \item Number of message-passing layers $L \in \{2,\dots, 6\}$
    \item Hidden layer dimension $d_h \in \{64, 128, 256, 512\}$
    \item Batch size $\in \{32, 64, 128, 256\}$
    \item Optimizer $\in \{\text{Adam, RMSProp, SGD}\}$
    \item Learning rate $\in [10^{-5}, 10^{-2}]$ (log scale)
    \item Learning rate scheduler $\in \{\text{none, cosine annealing, ReduceLROnPlateau}\}$
    \item Weight decay $\in [10^{-6}, 10^{-3}]$ (log scale)
\end{itemize}
  Stage 2 then tuned the physics-informed loss parameters on the best-performing model from Stage 1, namely the loss weight $\lambda \in [0, 0.5]$ and the warmup epoch at which physics-informed regularization is introduced $\in \{0, 10, 50, 100, 200\}$. As the physics-informed loss component is currently in a prototypical stage, this two-stage strategy was adopted to isolate its influence from the core architectural search and to allow independent refinement of each component.

To avoid computational waste on unpromising trials, a pruner was employed. To avoid discarding promising trials too early, the pruner had two grace periods. First, the initial five trials were always allowed to run to completion, giving the sampler enough data to make informed pruning decisions. Subsequently, each new trial was granted a 50-step warmup period during which no pruning could occur, ensuring that intermediate results were sufficiently accumulated before a trial could be terminated early. The study was distributed across four parallel workers on an A40 GPU node within the Lawrencium cluster. Each worker executed 15 trials in Stage 1 (60 total), with each trial permitted a maximum of 1,000 training epochs, followed by 9 trials per worker in Stage 2 (36 total).

\end{appendices}

\section*{Demo}
An interactive demonstration of SuperCond-GNN is available on Hugging Face Spaces, allowing users to specify circuit topologies and input parameters and obtain predicted nodal voltage distributions. The demo is accessible at {\url{https://huggingface.co/spaces/nmenon97/SuperCondGNN}}. Note that the hosted demo runs on Hugging Face's base CPU tier and will therefore exhibit longer inference times.

\bibliographystyle{ieeetr}  
\bibliography{references}

@techreport{berger_stability_2011,
	title = {Stability of superconducting cables with twisted stacked {YBCO} coated conductors},
	url = {https://dspace.mit.edu/handle/1721.1/93343},
	language = {en},
	urldate = {2026-04-16},
	institution = {MIT Plasma Science and Fusion Center},
	author = {Berger, A. D.},
	month = feb,
	year = {2011},
}

@article{yu_experimental_2024,
	title = {Experimental {Studies} on {Quench} {Behavior} {Measurements} of {HTS} {Tapes} {With} {Various} {Heater} {Configurations}},
	volume = {34},
	issn = {1558-2515},
	url = {https://ieeexplore.ieee.org/document/10387250},
	doi = {10.1109/TASC.2024.3351795},
	number = {5},
	urldate = {2026-06-02},
	journal = {IEEE Transactions on Applied Superconductivity},
	author = {Yu, Hui and Tang, Bohan and Yang, Shige and Jiang, Shili and Jiang, Donghui and Kuang, Guangli},
	month = aug,
	year = {2024},
	pages = {1--5},
}

@misc{vogt_ngspice_nodate,
	title = {Ngspice, the open source {Spice} circuit simulator - {Intro}},
	url = {https://ngspice.sourceforge.io/index.html},
	author = {Vogt, Holger},
}

@article{dezoort_graph_2023,
	title = {Graph neural networks at the {Large} {Hadron} {Collider}},
	volume = {5},
	copyright = {2023 Springer Nature Limited},
	issn = {2522-5820},
	url = {https://www.nature.com/articles/s42254-023-00569-0},
	doi = {10.1038/s42254-023-00569-0},
	language = {en},
	number = {5},
	urldate = {2026-05-22},
	journal = {Nature Reviews Physics},
	publisher = {Nature Publishing Group},
	author = {DeZoort, Gage and Battaglia, Peter W. and Biscarat, Catherine and Vlimant, Jean-Roch},
	month = may,
	year = {2023},
	pages = {281--303},
}

@article{marchevsky_quench_2021,
	title = {Quench {Detection} and {Protection} for {High}-{Temperature} {Superconductor} {Accelerator} {Magnets}},
	volume = {5},
	copyright = {http://creativecommons.org/licenses/by/3.0/},
	issn = {2410-390X},
	url = {https://www.mdpi.com/2410-390X/5/3/27},
	doi = {10.3390/instruments5030027},
	language = {en},
	number = {3},
	urldate = {2026-05-22},
	journal = {Instruments},
	publisher = {Multidisciplinary Digital Publishing Institute},
	author = {Marchevsky, Maxim},
	month = sep,
	year = {2021},
	pages = {27},
}

@article{rasheed_digital_2020,
	title = {Digital {Twin}: {Values}, {Challenges} and {Enablers} {From} a {Modeling} {Perspective}},
	volume = {8},
	issn = {2169-3536},
	shorttitle = {Digital {Twin}},
	url = {https://ieeexplore.ieee.org/document/8972429},
	doi = {10.1109/ACCESS.2020.2970143},
	urldate = {2026-05-22},
	journal = {IEEE Access},
	author = {Rasheed, Adil and San, Omer and Kvamsdal, Trond},
	year = {2020},
	pages = {21980--22012},
}

@inproceedings{lannutti_cuspice_2014,
	address = {Venice, Italy},
	title = {{CUSPICE} {The} revolutionary {NGSPICE} on {CUDA} {Platforms}},
	booktitle = {12th {MOS}-{AK} {Workshop} at the {ESSDERC}/{ESSCIRC} {Conference}},
	author = {Lannutti, Francesco and Menichelli, F. and Olivieri, M.},
	year = {2014},
}

@article{mo_graph_2022,
	title = {Graph {Theory}-{Based} {Programmable} {Topology} {Derivation} of {Multiport} {DC}–{DC} {Converters} {With} {Reduced} {Switches}},
	volume = {69},
	issn = {1557-9948},
	url = {https://ieeexplore.ieee.org/abstract/document/9467494},
	doi = {10.1109/TIE.2021.3090711},
	number = {6},
	urldate = {2026-05-14},
	journal = {IEEE Transactions on Industrial Electronics},
	author = {Mo, Liping and Chen, Guipeng and Huang, Jiangming and Qing, Xinlin and Hu, Yihua and He, Xiangning},
	month = jun,
	year = {2022},
	pages = {5745--5755},
}

@article{khan_weakly_2026,
	title = {A {Weakly} {Supervised} {Machine} {Learning} {Procedure} for {Acoustic} {Emission} {Quench} {Diagnostics}},
	volume = {36},
	doi = {10.1109/TASC.2025.3647009},
	number = {3},
	journal = {IEEE Transactions on Applied Superconductivity},
	author = {Khan, Maira and Krave, Steven and Marinozzi, Vittorio and Ngadiuba, Jennifer and Stoynev, Stoyan and Tran, Nhan},
	year = {2026},
	pages = {1--6},
}

@misc{hakhamaneshi_pretraining_2022,
	title = {Pretraining {Graph} {Neural} {Networks} for few-shot {Analog} {Circuit} {Modeling} and {Design}},
	url = {https://arxiv.org/abs/2203.15913v2},
	language = {en},
	urldate = {2026-05-14},
	journal = {arXiv.org},
	author = {Hakhamaneshi, Kourosh and Nassar, Marcel and Phielipp, Mariano and Abbeel, Pieter and Stojanović, Vladimir},
	month = mar,
	year = {2022},
}

@misc{xiao_surrogate_2026,
	title = {A {Surrogate} model for {High} {Temperature} {Superconducting} {Magnets} to {Predict} {Current} {Distribution} with {Neural} {Network}},
	url = {https://arxiv.org/abs/2509.06067},
	author = {Xiao, Mianjun and Song, Peng and Liu, Yulong and Korte, Cedric and Xu, Ziyang and Gao, Jiale and Lu, Jiaqi and Nie, Haoyang and Deng, Qiantong and Qu, Timing},
	year = {2026},
	note = {\_eprint: 2509.06067},
}

@article{khamis_circuit_2024,
	title = {Circuit topology aware {GNN}-based multi-variable model for {DC}-{DC} converters dynamics prediction in {CCM} and {DCM}},
	volume = {36},
	issn = {1433-3058},
	url = {https://doi.org/10.1007/s00521-024-10293-0},
	doi = {10.1007/s00521-024-10293-0},
	language = {en},
	number = {33},
	urldate = {2026-05-14},
	journal = {Neural Computing and Applications},
	author = {Khamis, Ahmed K. and Agamy, Mohammed},
	month = nov,
	year = {2024},
	pages = {20807--20822},
}

@misc{akiba_optuna_2019,
	title = {Optuna: {A} {Next}-generation {Hyperparameter} {Optimization} {Framework}},
	shorttitle = {Optuna},
	url = {https://arxiv.org/abs/1907.10902v1},
	language = {en},
	urldate = {2026-05-05},
	journal = {arXiv.org},
	author = {Akiba, Takuya and Sano, Shotaro and Yanase, Toshihiko and Ohta, Takeru and Koyama, Masanori},
	month = jul,
	year = {2019},
}

@article{sakakibara_experimental_2026,
	title = {Experimental {Investigation} of {CNN}-{Based} {Voltage} {Prediction} for {REBCO} {Pancake} {Coil} {Protection}},
	volume = {36},
	issn = {1558-2515},
	url = {https://ieeexplore.ieee.org/abstract/document/11474898},
	doi = {10.1109/TASC.2026.3680862},
	number = {5},
	urldate = {2026-04-20},
	journal = {IEEE Transactions on Applied Superconductivity},
	author = {Sakakibara, Riki and Mato, Takanobu and Inoue, Ryota and Ueda, Hiroshi and Kim, SeokBeom and Noguchi, So},
	month = aug,
	year = {2026},
	pages = {1--5},
}

@misc{pfaff_learning_2021,
	title = {Learning {Mesh}-{Based} {Simulation} with {Graph} {Networks}},
	url = {http://arxiv.org/abs/2010.03409},
	doi = {10.48550/arXiv.2010.03409},
	urldate = {2026-04-16},
	publisher = {arXiv},
	author = {Pfaff, Tobias and Fortunato, Meire and Sanchez-Gonzalez, Alvaro and Battaglia, Peter W.},
	month = jun,
	year = {2021},
	note = {arXiv:2010.03409 [cs]},
}

@article{teyber_numerical_2022,
	title = {Numerical investigation of current distributions around defects in high temperature superconducting {CORC}® cables},
	volume = {35},
	issn = {0953-2048},
	url = {https://doi.org/10.1088/1361-6668/ac86fd},
	doi = {10.1088/1361-6668/ac86fd},
	language = {en},
	number = {9},
	urldate = {2026-04-16},
	journal = {Superconductor Science and Technology},
	publisher = {IOP Publishing},
	author = {Teyber, Reed and Marchevsky, Maxim and Martinez, Aurora Cecilia Araujo and Prestemon, Soren and Weiss, Jeremy and van der Laan, Danko},
	month = aug,
	year = {2022},
	pages = {094008},
}

@article{martinez_electric-circuit_2020,
	title = {An {Electric}-{Circuit} {Model} on the {Inter}-{Tape} {Contact} {Resistance} and {Current} {Sharing} for {REBCO} {Cable} and {Magnet} {Applications}},
	volume = {30},
	issn = {1558-2515},
	url = {https://ieeexplore.ieee.org/document/8986625},
	doi = {10.1109/TASC.2020.2972215},
	number = {4},
	urldate = {2026-04-16},
	journal = {IEEE Transactions on Applied Superconductivity},
	author = {Martínez, Aurora Cecilia Araujo and Ji, Qing and Prestemon, Soren O. and Wang, Xiaorong and Maury Cuna, Georfrey Humberto I.},
	month = jun,
	year = {2020},
	pages = {1--5},
}

@article{kang_current_2023,
	title = {The {Current} {Unbalance} in {Stacked} {REBCO} {Tapes} — {Simulations} {Based} on a {Circuit} {Grid} {Model}},
	volume = {33},
	issn = {1558-2515},
	url = {https://ieeexplore.ieee.org/document/10286421},
	doi = {10.1109/TASC.2023.3324990},
	number = {9},
	urldate = {2026-04-16},
	journal = {IEEE Transactions on Applied Superconductivity},
	author = {Kang, Rui and Wang, Juan and Feng, Ze and Xu, Qingjin},
	month = dec,
	year = {2023},
	pages = {1--12},
}

@article{willering_effect_2015,
	title = {Effect of variations in terminal contact resistances on the current distribution in high-temperature superconducting cables},
	volume = {28},
	issn = {0953-2048},
	url = {https://doi.org/10.1088/0953-2048/28/3/035001},
	doi = {10.1088/0953-2048/28/3/035001},
	language = {en},
	number = {3},
	urldate = {2026-04-16},
	journal = {Superconductor Science and Technology},
	publisher = {IOP Publishing},
	author = {Willering, G P and van der Laan, D C and Weijers, H W and Noyes, P D and Miller, G E and Viouchkov, Y},
	month = jan,
	year = {2015},
	pages = {035001},
}

@article{pothavajhala_experimental_2014,
	title = {Experimental and {Model} {Based} {Studies} on {Current} {Distribution} in {Superconducting} {DC} {Cables}},
	volume = {24},
	issn = {1558-2515},
	url = {https://ieeexplore.ieee.org/abstract/document/6603254},
	doi = {10.1109/TASC.2013.2282568},
	number = {3},
	urldate = {2026-04-16},
	journal = {IEEE Transactions on Applied Superconductivity},
	author = {Pothavajhala, Venkata and Graber, Lukas and Kim, Chul Han and Pamidi, Sastry},
	month = jun,
	year = {2014},
	pages = {1--5},
}

\end{document}